\documentclass{article}

\usepackage{dsfont}
\usepackage{color}
\usepackage{graphicx, nicefrac}
\usepackage{multicol}
\usepackage{multirow,makecell, rotating}
\usepackage[table,xcdraw]{xcolor}
\usepackage{makecell}
\usepackage{acronym}
\usepackage{pifont}
\graphicspath{ {./LaTeX/figures }}
\usepackage[breaklinks]{hyperref}
\hypersetup{
    colorlinks=true,
    urlcolor=blue,
    }
\usepackage{spconf,amsmath,graphicx}
\usepackage{subcaption}
\usepackage[font=small, labelfont=bf, skip=5pt]{caption}

\newcommand{\pmr}[1]{\scriptsize$\pm$#1}

\acrodef{DSP}{Digital Signal Processing}
\acrodef{SSR}{Speech Super Resolution}
\acrodef{BWE}{Bandwidth Extension}
\acrodef{LSD}{Log Spectral Distance}
\acrodef{STFT}{short-time Fourier transform}
\acrodef{iSTFT}{inverse short-time Fourier transform}
\acrodef{MSD}{Multi-Scale Discriminators}
\acrodef{FTB}{Frequency Transformer Block}

\newcommand{\moshe}[1]{{\color{purple}M: #1}}
\newcommand{\ortal}[1]{{\color{green}O: #1}}
\newcommand{\todo}[1]{{\color{blue}TODO: #1}}

\usepackage{bm}
\usepackage{amssymb,amsmath,graphicx,bbm,color,booktabs}
\usepackage{amsopn}

\newcommand{\vx}{\bm{x}}

\newcommand{\R}{\mathbb{R}}

\renewcommand{\eqref}[1]{Eq.~(\ref{#1})}

\title{Aero: Audio Super Resolution in the Spectral Domain}
\name{Moshe Mandel, Or Tal, Yossi Adi}
\address{School of Computer Science and Engineering \\ The Hebrew University of Jerusalem, Israel}

\begin{document}

\maketitle

\begin{abstract}
We present AERO, a audio super-resolution model that processes speech and music signals in the spectral domain. AERO is based on an encoder-decoder architecture with U-Net like skip connections. We optimize the model using both time and frequency domain loss functions. Specifically, we consider a set of reconstruction losses together with perceptual ones in the form of adversarial and feature discriminator loss functions. To better handle phase information the proposed method operates over the complex-valued spectrogram using two separate channels. Unlike prior work which mainly considers low and high frequency concatenation for audio super-resolution, the proposed method directly predicts the full frequency range. We demonstrate high performance across a wide range of sample rates considering both speech and music. AERO outperforms the evaluated baselines considering Log-Spectral Distance, ViSQOL, and the subjective MUSHRA test. Audio samples and code are available \href{https://pages.cs.huji.ac.il/adiyoss-lab/aero/}{here}.
\end{abstract}

\vspace{-0.2cm}
\section{Introduction}
\label{sec:intro}

% describe the method in more details than the abstract.
% \ortal{try to move figure to next page so it won't interrupt the introduction flow}

% What is the problem?
Audio super-resolution, also referred to as \ac{BWE}, is the task of generating an audio waveform at a higher sampling rate from its corresponding low-resolution signal~\cite{springer}. High resolution audio contains greater detail, resulting in a crispier and a more natural sound overall.

% Why is it interesting and important?
% High resolution audio contains more detail, is richer, crispier and more natural.
% telecommunication systems transfer at a low resolution.
% generative deep learning methods generate at low resolution.
% Post processing tools operate on higher resolution signals (is this true?).
% Thus, BWE is an important task for audio/video calls or any transmission system, and any generative pipeline.

Due to hardware or transmission limitations, it is common that a speech signal's frequency bandwidth is reduced when transmitted via telecommunication systems. Furthermore, most deep learning based methods for generating audio, both in speech and music domains, are limited to a target audio of low resolution. For instance, in~\cite{audiogen} audio is generated from descriptive text captions at a target rate of 16~kHz. This results in losing most of the signal's richness and fidelity, yielding poor user experience. As many audio applications and post-processing tools today benefit from using the full frequency bandwidth, \ac{BWE} is an important task in itself for audio and video calls, and can also be beneficial for generative audio systems and post-processing pipelines.

% Why is it hard? (E.g., why do naive approaches fail?)
% Over the last decades, much reseach has shown success on this task.
% Early methods were based on a source-filter model of speech production.
% Deep learning methods that use reconstuction loss only, adverasarial methods, diffusion methods.
% here starts the related work.
Over the last decades, a plethora of research showed success in \ac{BWE}. Early works base their solutions on the source-filter model~\cite{springer}. Generative methods that utilize neural networks provide statistical and data-driven solutions that excel in generating high-frequency signals. Early methods are trained by minimizing a reconstruction loss~\cite{li, kuleshov}. Later, generative adversarial networks, diffusion based methods and other generative models showed further success~\cite{seanet, mugan, bwe-all-you-need, nuwave2, wsrglow}.

\begin{figure}[t!]
    \centering
    \includegraphics[width=1\linewidth]{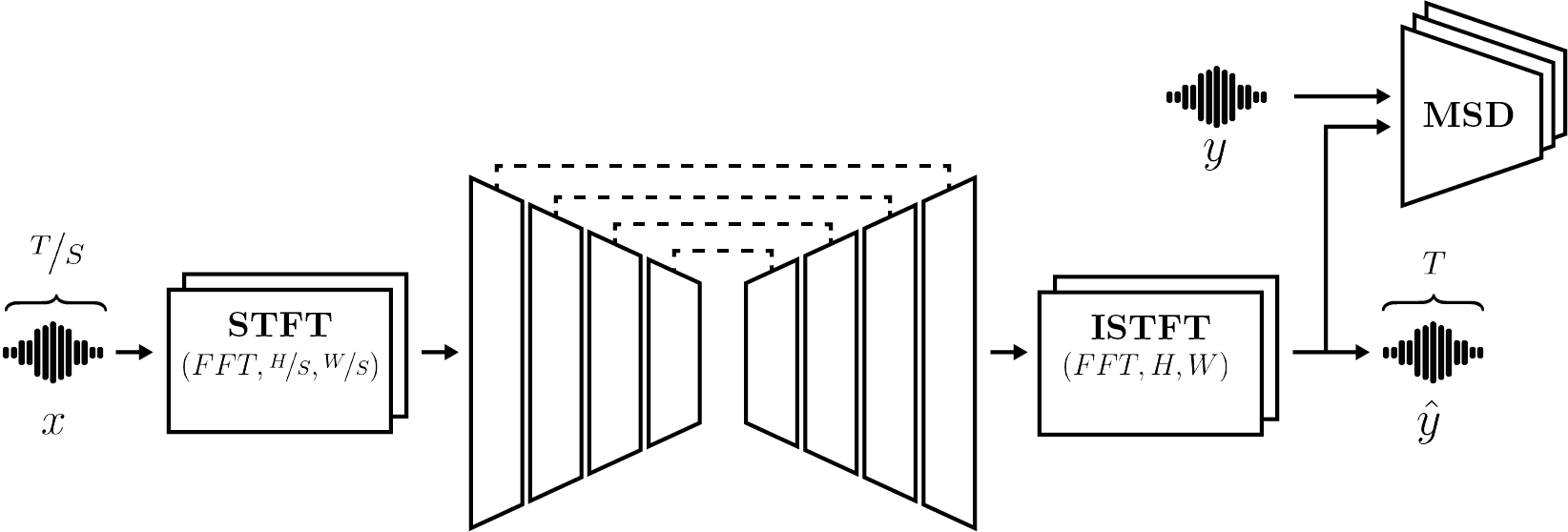}
    \caption{A waveform of length $\nicefrac{T}{S}$ is converted to the complex-as-channels spectrogram via STFT with scaled down hop and window sizes: $\nicefrac{H}{S}, \nicefrac{W}{S}$. After passing through a U-Net-like model, the signal is converted back to a waveform of length $T$ via iSTFT with hop and window sizes: $H, W$. During training, a set of multi-scale discriminators (MSD) are utilized for adversarial and feature losses in the time domain, together with a spectral reconstruction loss.\label{fig:model}}    
    \vspace{-0.2cm}
\end{figure}

% Why hasn't it been solved before? (Or, what's wrong with previous proposed solutions? How does mine differ?)
Neural network based methods either operate on the raw waveform~\cite{seanet, kuleshov, mugan, bwe-all-you-need, nuwave2} or on some spectral representation of the signal: Line Spectral Frequencies (LSF)~\cite{lsf}, magnitude spectrogram~\cite{nugan} or the log-power magnitude spectrogram~\cite{li, eskimez}.
The length of audio signals, especially at high resolution, is extremely long, hence modeling it is computationally expensive; this is especially relevant in time-domain based methods. Spectral-based methods face the following challenges: (i) utilizing the phase information from the input, and (ii) reusing the low-frequency bands of the input signal in the reconstructed signal.
The authors in \cite{li, eskimez} resolve (i) by flipping the existing low-resolution phase, whereas \cite{phase-aware} applies a designated neural network to directly estimate the high-frequency phase. Typical spectral-based approaches~\cite{li, eskimez, nugan, phase-aware} resolve (ii) by generating only the high-frequency spectral features and reusing the low-frequency features via concatenation. This may cause artifacts at the verge between existing and generated frequency bands. 

% % What are the key components of my approach and results? Also include any specific limitations.
% Demucs is a convolutional U-net model that has demonstrated great performance on various speech-related tasks: Source separation\cite{hdemucs}, and real-time speech signal enhancement \cite{defossez2020real}. \cite{audioPrior} has demonstrated that the model contains inherent priors that allow it to perform well in general audio-related tasks.

In this work we propose a method for generating high-frequency content in the spectral domain. Inspired by recent success of the Demucs architecture on music source separation and speech enhancement~\cite{hdemucs,defossez2020real,audioPrior}, the proposed method is composed of a convolutional U-Net model that operates solely in the frequency domain, together with a set of reconstruction, adversarial and feature losses that operate on the spectral and time representations of the signal. Similar to \cite{behmgan}, the model operates on the Complex-as-Channel representation of the complex-valued spectrogram \cite{choi}, thus jointly utilizing both magnitude and phase information and avoiding the need to separately reconstruct the phase of the high-frequency signal. Furthermore, we introduce a way to upsample a signal in the spectral domain that avoids concatenation between existing and generated frequency bands. The proposed method is illustrated in Figure~\ref{fig:model}.

% Summary of Contributions
We empirically show that the proposed method surpasses current state-of-the-art methods. We perform ablation studies to investigate how objective and subjective metrics are impacted by different components of the U-Net model and by the various losses used.

\begin{figure}[t!]
    \centering
    \includegraphics[width=\linewidth]{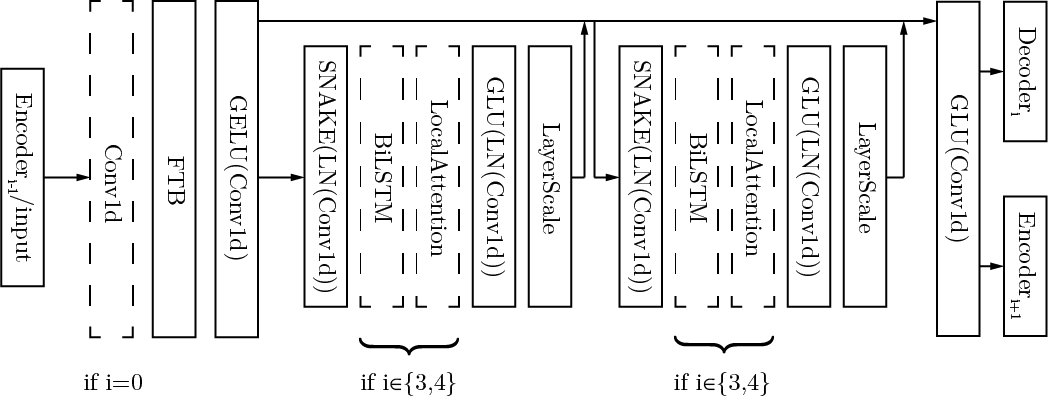}
    \caption{Encoder layer.}
    \label{fig:encoder-block}
\end{figure}

\vspace{-0.2cm}
\section{Method}
\label{sec:model}

% \begin{figure*}[t!]
%     \centering
%     \begin{subfigure}[]{0.8\textwidth}
%         \includegraphics[width=\textwidth]{graphics/encoder-decoder-bg-2.png}
%         \caption{A waveform of length $\nicefrac{T}{S}$ is converted to the complex-as-channels spectrogram via STFT with scaled down hop and window sizes: $\nicefrac{H}{S}, \nicefrac{W}{S}$. After passing through a U-net-like model, the signal is converted back to a waveform of length $T$ via ISTFT with hop and window sizes: $H, W$.}
%         \label{fig:model}
%     \end{subfigure}%
    
%     \begin{subfigure}[]{0.8\textwidth}
%         \includegraphics[width=0.9\linewidth]{graphics/encoder-block-abbriviated-rotated.png}
%         \caption{Encoder block.}
%         \label{fig:encoder-block}
%     \end{subfigure}
%     \caption{Model architecture.}
%     \label{fig:architecture}
% \end{figure*}

Given a signal of low resolution $x{\in}\R^{\nicefrac{T}{s}}$ that was downsampled from its high-resolution counterpart $y{\in}\R^{T}$, where $s$ is a fixed upsampling scaling factor, our goal is to reconstruct $\hat{y}\approx{y}$ and generate its missing high-frequency content.

Operating in the frequency domain, the waveform signal $x$ is converted to $X$ using \ac{STFT}. As demonstrated in \cite{choi}, the input to the model is represented as a concatenation of the real and imaginary parts of the complex-valued spectrogram. The model is then trained to directly predict the spectrogram of the high-frequency signal. The resulting spectrogram $\hat{Y}$ is then transformed back to the reconstructed high-resolution signal $\hat{y}$ using the \ac{iSTFT}.
\vspace{-0.2cm}
\subsection{Architecture}

Inspired by Demucs \cite{hdemucs}, our model is a convolutional U-Net architecture that operates in the frequency domain, with four layers in the encoder and decoder each. 
% Detailed in Figure~\ref{fig:encoder-block}, t
The encoder accepts the signal in spectrogram form, and uses 1D convolutions that operate only on the frequency axis. A \ac{FTB} \cite{phasen} is added before each encoder layer. Within each layer are two compressed residual branches; as opposed to the original Demucs architecture, we use the Snake activation function \cite{snake}. For the inner layers of the encoder we utilize both LSTM and temporal-based attention modules. With a downsampling scheme of [4,4,2,2], the resulting latent vector is a projection of the input spectrogram, compressed $64$-fold in the frequency axis. The encoder layer is visually described in Figure~\ref{fig:encoder-block}. Following the encoder, a decoder transforms the latent vector to a spectrogram of size equal to that of the input to the encoder.

Unlike prior studies that first perform upsampling and then optimize the network to fill the missing frequency range\cite{seanet, nugan}, in this work we directly multiply the hop-length and the window size of the \ac{iSTFT} by the scaling factor $s$ (more details can be found in Section~\ref{subsec:spec_upsampling}). With the proposed upsampling technique, information in the encoding process is held across the whole range of frequencies instead of being limited by the Nyquist rate. To accommodate for this, we use concatenated skip connections instead of summation.

\begin{table*}[t!]
\centering
\caption{L, V and M denote LSD, ViSQOL and MUSHRA respectively. MUSHRA score is specified with a $\pm$ Confidence Interval of 0.95. \label{tab:results}}
\resizebox{\textwidth}{!}{%
\begin{tabular}{@{\hskip6pt}l|ccc|ccc|ccc|ccc} 
    \toprule
     & 
     \multicolumn{3}{|c|}{8-16} & 
     \multicolumn{3}{|c|}{8-24} & 
     \multicolumn{3}{|c|}{4-16} & 
     \multicolumn{3}{|c}{11-44} \\
    \midrule
     & L$\downarrow$ & V$\uparrow$ & M$\uparrow$ \
     & L$\downarrow$ & V$\uparrow$ & M$\uparrow$ \
     & L$\downarrow$ & V$\uparrow$ & M$\uparrow$ \
     & L$\downarrow$ & V$\uparrow$ & M$\uparrow$ \\
    \midrule
        Reference &- &- &96.25\pmr{1.5} &- &- &97.16\pmr{1.4} &- &- &96.18\pmr{1.5} &- &- &95.30\pmr{2.5} \\
        Anchor &- &- &54.65\pmr{4.3} &- &- &56.21\pmr{4.4} &- &- &41.14\pmr{3.8} &- &- &46.55\pmr{7.4} \\
        \midrule
        Sinc &2.32 &3.41 &60.13\pmr{4.7} &2.96 &3.41 &59.49\pmr{4.8} &3.59 &2.27 &43.03\pmr{3.9} &3.91 &1.97 &47.61\pmr{8.0} \\
        TFiLM \cite{kuleshov} &1.27 &3.18 &58.53\pmr{4.0} &- &- &- &1.77 &2.25 &41.91\pmr{4.0} &- &- &- \\
        SEANet \cite{seanet} &0.79 &4.08 &91.23\pmr{2.9} &0.91 &4.06 &94.16\pmr{2.2} &0.99 &3.16 &89.40\pmr{3.2} &1.13 &2.88 &80.52\pmr{7.0} \\
        BEHMGAN \cite{behmgan} &- &- &- &- &- &- &- &- &- &1.80 &2.01 &46.27\pmr{8.3} \\
        \midrule
        Ours ($\nicefrac{256}{512}$) &0.84 &4.02 &90.58\pmr{2.3} &0.99 &4.03 &\textbf{96.40\pmr{1.9}} &1.04 &3.04 &86.14\pmr{3.4} &1.16 &2.88 &81.21\pmr{6.4} \\
        Ours ($\nicefrac{128}{512}$) &0.80 &4.11 &92.63\pmr{2.4} &0.91 &4.12 &95.41\pmr{2.0} &0.99 &3.15 &\textbf{92.05\pmr{2.7}} &1.16 &\textbf{2.89} &81.67\pmr{6.8} \\
        Ours ($\nicefrac{64}{512}$) &\textbf{0.77} &\textbf{4.16} &\textbf{94.64\pmr{1.6}} &\textbf{0.90} &\textbf{4.17} &94.45\pmr{2.1} &\textbf{0.94} &\textbf{3.28} &90.61\pmr{3.1} &\textbf{1.12} &2.88 &\textbf{84.18\pmr{5.6}} \\
    \bottomrule
\end{tabular}
}
\end{table*}

\begin{table}[t!]
\small
\centering
\caption{12-48 results. We use the $\nicefrac{256}{1024}$ configuration.\label{tab:12-48-results}}

\begin{tabular}{@{\hskip6pt}l|ccc} 
    \toprule
     & L$\downarrow$ & V$\uparrow$ & M$\uparrow$ \\     
    \midrule
        Reference &- &- &98.47\pmr{0.9} \\
        Anchor &- &- &67.76\pmr{4.1} \\
        \midrule
        Sinc &3.36 &4.33 &69.77\pmr{4.3} \\
        SEANet \cite{seanet} &\textbf{0.86} &\textbf{4.71} &96.17\pmr{1.6} \\
        NuWave2 \cite{nuwave2} &1.34 &4.42 &84.87\pmr{4.5} \\
        \midrule
        Ours &0.92 &4.67 &\textbf{96.71\pmr{1.8}} \\
    \bottomrule
\end{tabular}

\end{table}

\begin{figure}[t!]
    \centering
    \begin{subfigure}{0.49\columnwidth}
        \centering
        \includegraphics[width=\linewidth]{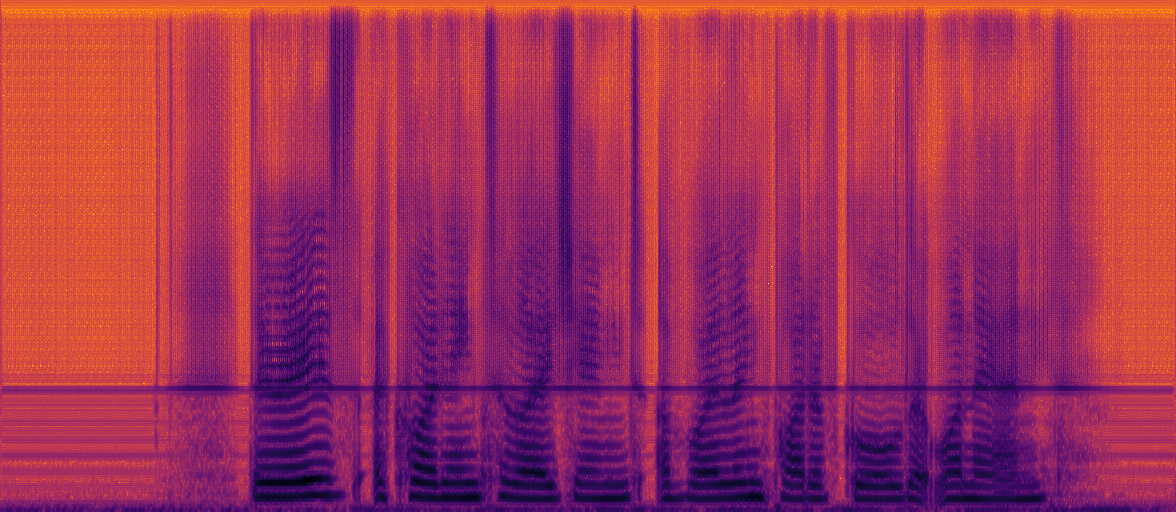}
        \label{fig:verge-artifact}
    \end{subfigure}%
    ~
    \begin{subfigure}{0.49\columnwidth}
        \centering
        \includegraphics[width=\linewidth]{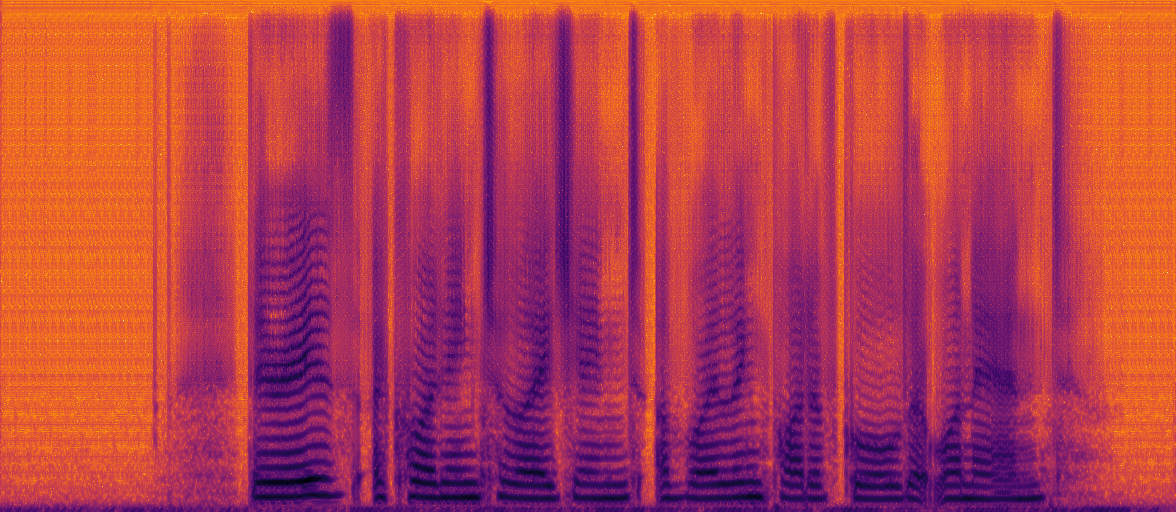}
        \label{fig:verge-no-artifact}
    \end{subfigure}
    \vspace{-0.4cm}
    \caption{Spectral upsampling comparison. Both spectrograms are outputs of the same signal resulting from the model before the iSTFT stage, upsampled from 4kHz to 16kHz. The left spectrogram is generated with no spectral upsampling, whereas the one on the right is generated with spectral upsampling.\label{fig:spectrograms}}
    \vspace{-0.2cm}
\end{figure}

\vspace{-0.2cm}
\subsection{Training Objective}
\sloppy Similar to ~\cite{defossez2020real}, we use a multi-resolution \ac{STFT} loss \cite{yamamoto} using FFT bins ${\in}\{512,1024,2048\}$, hop lengths ${\in}\{50,120,240\}$ and window sizes ${\in}\{240,600,1200\}$. 
Additionally we apply the multi-scale adversarial and feature losses~\cite{seanet} that operate in the time domain.

Using only a spectral reconstruction loss we substantially outperform state-of-the-art methods in objective metrics. We found that this produces audible artifacts and significantly impacts subjective metrics. Adding adversarial and feature losses acts as a type of perceptual loss and removes artifacts while slightly reducing the objective metrics (see Table~\ref{tab:adv_ablation}).

\vspace{-0.2cm}
\subsection{Upsampling in the Frequency Domain}
\label{subsec:spec_upsampling}

As mentioned above, a typical first step in audio upsampling methods is to upsample the input signal to the target sample rate in the time domain using Sinc interpolation \cite{tfilm, nugan, nuwave2, seanet, bwe-all-you-need, mugan}. The resulting waveform potentially holds the Nyquist rate of a high-resolution signal, but, as no additional frequencies are actually added, the top segment of the corresponding spectrogram holds no significant information. We find that this technique produces significant artifacts at the verge between the observed low-frequency range and the generated high-frequency range, as seen in Figure~\ref{fig:spectrograms}.

% Noted. commenting out for now for clarity in pdf 
% \ortal{the upsampling method is a bit unclear, I think you should simplify things, maybe add a scheme or specify the iterative algorithm to make things simpler here, I didn't get it.\\
% As this section is your suggested novelty in the paper I think you could give it it's own section, and explain it in the most simple manner you can}

To mitigate this, we propose a method for upsampling in the frequency domain. Using different window sizes and hop lengths at the \ac{STFT} and \ac{iSTFT} stages, we can start from a given low-resolution signal $\vx \in \R^{\nicefrac{T}{s}}$ and end with a high-resolution $y \in \R^{T}$ -- upsampled by a factor of $s$ -- while using a single \ac{STFT} representation of fixed size at the intermediate generation stage. With this technique, the input to the model holds information across the whole range of frequencies.

At the \ac{STFT} stage, the waveform signal $x$ is converted to $X$ with $f$ frequency bins, a hop length of $\frac{\nicefrac{f}{k}}{s}$, and a window length of $\frac{f}{s}$, where $k$ defines the overlapping ratio between consecutive frames.
At the \ac{iSTFT} stage, the model's output $\hat{Y}$ is transformed to $\hat{y}$ using parameters customized to the upsampling setting; this uses the same number of frequency bins $f$, but with a hop length of $\nicefrac{f}{k}$ and a window length of $f$. This process is illustrated in Figure~\ref{fig:model}.
\vspace{-0.2cm}
\section{Experiments}
\label{sec:experiments}

We test the performance of our model on speech signals from VCTK \cite{vctk} under different upsampling settings: 4-16~kHz, 8-16~kHz, 8-24~kHz, and 12-48~kHz. 
In addition to speech signals, we test our model on music signals from MusDB \cite{musdb18-hq} in a setting of 11.025-44.1 kHz.
We measure our model's performance using both objective and qualitative measures.

% As noted by \cite{nuwave2} \todo{find citation}, SiSNR is not a suitable metric for the task of \ac{SSR}.
% \subsection{Data}
The VCTK dataset \cite{vctk} contains around 44 hours of speech from 110 speakers, sampled at 48~kHz. The training and test setup is the same as in \cite{nuwave2}: we omit speakers $p280$ and $p315$. Using only recordings from $mic1$, we use the first 100 speakers for training and the remaining 8 speakers for testing.

The MusDB dataset \cite{musdb18-hq} contains 150 songs (10 hours) of musical mixtures, along with isolated stems, sampled at 44.1~kHz. We use the provided train/test setup, using only the mixture tracks to test our model and baselines.
\vspace{-0.2cm}

\subsection{Baselines}
% In all settings and data-sets, we use the default \code{rate} command of the Sox package for resampling the data.
We compare the proposed method to several audio super resolution methods. For speech data, we compare against SEANet~\cite{seanet}, TFiLM~\cite{kuleshov}, and NuWave2~\cite{nuwave2}. For musical data we compare against SEANet and BEHM-GAN~\cite{behmgan}. We additionally include a comparison to a naive Sinc interpolation method using the \texttt{torchaudio} package. 

% In all settings and data-sets, we compare the proposed method to the Seanet model \cite{seanet}. For the 8-16 kHz and 4-16 kHz settings, we use the official implementation of TFiLM \cite{kuleshov}, trained on the VCTK dataset. For the 12-48 kHz setting, we use the provided pre-trained model of NuWave2 \cite{nuwave2}, to produce the results. For all settings, we use the sinc-interpolation method provided by the \code{torchaudio.transforms.Resample} function.

% For comparison in the musical domain, in addition to SEANet -- we use the official implementation of BEHM-GAN \cite{behmgan}, and retrain its bandwidth-extension model on the MUSDB dataset.
% \vspace{-0.2cm}
\subsection{Model Evaluation}
Two objective metrics and a qualitative metric are used to measure the quality of the reconstructed audio with respect to the reference signal:

\noindent {\bf \ac{LSD}} denoting the log-spectral power magnitudes $\hat{Y}$ and $Y$ of signals $\hat{y}$ and $y$, defined as $Y(\tau, \kappa)=\log_{10}|S(y)|^2$ where $S$ is the \ac{STFT}, with the Hanning window of $2048$ samples and hop length of $512$. The \ac{LSD} is defined as:     
    \[ LSD(\hat{y},y)=\frac{1}{T}\sum_{\tau=1}^{T} \sqrt{ \frac{1}{K} \sum_{\kappa=1}^{K} (\hat{Y}(\tau, \kappa) - Y(\tau, \kappa) )^2  }. \]
    
\noindent {\bf Virtual Speech Quality Objective Listener (ViSQOL)} is a signal-based, full-reference, intrusive metric that models human speech quality perception using a spectro-temporal measure of similarity between a reference and a test speech signal (from 1 to 5)~\cite{visqol}. The MusDB setting was evaluated using the audio mode configuration.

\noindent {\bf MUlti Stimulus test with Hidden Reference and Anchor (MUSHRA)} is a qualitative subjective assessment of intermediate quality level of audio systems. We used a web platform \cite{webmushra} to perform a human listening test, asking users to rate the quality of recordings in the range of 0 to 100~\cite{mushra}.    
% \vspace{-0.2cm}

\subsection{Results}
In all experiments, we trained our model for $500K$ steps. For the baseline methods, we followed the recommended hyper-parameters recipes provided by the authors.

In settings with a target sampling frequency of up to 44.1~kHz, we use a batch-size of 16, $\frac{\text{hop length}}{\text{window size}}$ ratios of $k{\in}\{\frac{1}{2},\frac{1}{4},\frac{1}{8}\}$ and 512 frequency bins. For the 12-48~kHz setting, we use a batch-size of 8, $k{=}\frac{1}{4}$, and 1024 frequency bins. 

Results are summarized in Tables~\ref{tab:results} and~\ref{tab:12-48-results}. The results suggest that the proposed method is superior to the evaluated baselines considering both objective and subjective metrics. Note that under the 4-16~kHz setting, we outperform all methods by a significant margin with $\frac{\text{hop length}}{\text{window size}}$ ratio of  $k{=}\frac{1}{8}$, with $k{=}\frac{1}{4}$ we are comparable with the adversarial method SEANet in objective metrics, while surpassing it subjectively, whereas with $k=\frac{1}{2}$ we are surpassed by it.
% \vspace{-0.2cm}
\subsection{Ablation Study}
\label{subsec:ablation}
We study the effect of the different discriminators and encoder components on the overall model performance. 

\begin{table}[!t]
\small
\centering
\caption{Adversarial ablation study \label{tab:adv_ablation}}
% \resizebox{0.9\columnwidth}{!}{%
\begin{tabular}{@{\hskip6pt}l|ccc} 
    \toprule
    Setting & LSD $\downarrow$ & VISQOL $\uparrow$ & MUSHRA $\uparrow$ \\
    \midrule
    Reference &- &- &92.49\pmr{2.2} \\
    Anchor &- &- &32.34\pmr{3.3} \\
    \cmidrule{1-4}
    No disc. &\textbf{0.8793} &\textbf{3.363} &32.46\pmr{3.5} \\
    1 MSD &0.978 &3.202 &\textbf{85.79\pmr{3.0}} \\
    3 MSD &0.943 &3.275 &85.57\pmr{2.9} \\
    Only feat. loss &0.986 &3.253 &77.64\pmr{3.7} \\
    Only adv. loss &1.012 &3.018 &73.96\pmr{4.0} \\
    \bottomrule
    \end{tabular}
% }
% \vspace{-0.2cm}
\end{table}

\noindent {\bf Impact of Discriminators.}
We investigate the usage of different number of \ac{MSD}, adversarial and feature losses, both jointly and each separately, and how they impact objective and subjective metrics. Results are reported in Table~\ref{tab:adv_ablation}. We note that while the non-adversarial setting provides the best objective results, it significantly under-performs in the subjective tests. The model with three \ac{MSD} with a combination of both adversarial and feature losses ranked second in objective metrics, but is slightly surpassed by the similar single discriminator setting in the subjective tests. We conclude that both feature and adversarial losses are crucial to the success of the task, whereas the exact number of \ac{MSD} used is less crucial.

\begin{table}[h!]
\small
\centering
\caption{Component ablation study. We report LSD and ViSQOL results for different model configuration. Under the Upsampling \mbox{column} "spec." denotes the proposed method and "time" denotes a pre-upsampling stage via Sinc interpolation. \label{tab:ablation}}
% \resizebox{1.0\columnwidth}{!}{%
\begin{tabular}{c|ccc|ccc} 
    \toprule
    & Activation & Upsampling & FTB & LSD  $\downarrow$ & VISQOL $\uparrow$ \\
    \midrule
    1 & ReLU & spec. & yes & 0.945 & 3.262 \\
    2 & ReLU & spec. & no & 0.952 & 3.273 \\
    3 & ReLU & time & yes & 0.957 & 3.263 \\
    4 & ReLU & time & no & 0.948 & 3.249 \\
    5 & Snake & spec. & yes & \bf 0.943 & \bf 3.275 \\
    6 & Snake & spec. & no & 0.958 & 3.243 \\
    7 & Snake & time & yes & 0.947 & 3.267 \\
    8 & Snake & time & no & 0.977 & 3.245 \\
    \bottomrule
\end{tabular}
% }
% \vspace{-0.2cm}
\end{table}

\noindent {\bf Component Analysis.}
We investigate the choice of activation function, upsampling technique, the usage of the \ac{FTB} module and their impact on the objective metrics. Results are reported in Table~\ref{tab:ablation}. While the study shows no significant improvement provided by individual components, the overall contribution of all the components provides the optimal results. As seen in Figure~\ref{fig:spectrograms}, using spectral upsampling produces a finer transition from the observed and generated frequency ranges.

\noindent {\bf Impact of Hop Length Size.} The choice of $\frac{\text{hop length}}{\text{window size}}$ for $k{\in}\{\frac{1}{2},\frac{1}{4},\frac{1}{8}\}$ defines the length of the signal in the spectral domain, which impacts the computational cost during training and inference. We tested training times on the VCTK's 8-16~kHz setting, and inference time on 5 songs from the musical 11-44~kHz setting, on a single Nvidia A5000 GPU. Comparing to $k{=}\frac{1}{8}$, we record an improvement by a factor of $1.4$ and $2.03$ in average training time per epoch, and $3.35$ and $8.43$ in average inference time, for $k{=}\frac{1}{4},\frac{1}{2}$ respectively. As observed in Table~\ref{tab:results}, we see a decline in objective metrics as $k$ increases. For every $k$, under all settings, our method is subjectively comparable or improves, with respect to the best evaluated baseline -- SEANet.
\section{Conclusions \& Future Work}
\label{sec:conclusions}
In this work we proposed an encoder-decoder method operating over the complex-valued spectrogram for audio super-resolution. We evaluate the proposed method using different upsampling factors considering both speech and music data. We empirically show that the proposed method is superior to the evaluated baselines considering objective and subjective metrics, and we conclude with an ablation study to better assess the contribution of each of the models' components to the model performance. 

For future work, we would like to explore the possibility to convert the proposed method to real-time and streaming processing. Additionally, we would like to train such models considering multi-task setups, e.g., jointly upsampling and denoising the input signal. 

\section*{Acknowledgements}
This research is supported by the \emph{science accelerator} program by the Ministry of Science and Technology, Israel.

\clearpage

\bibliographystyle{IEEEbib}
\bibliography{refs}

\begin{thebibliography}{10}

\bibitem{springer}
Bernd Iser et~al.,
\newblock {\em Bandwidth extension of speech signals},
\newblock Springer, 2008.

\bibitem{audiogen}
Felix Kreuk et~al.,
\newblock ``Audiogen: Textually guided audio generation,''
\newblock 2022.

\bibitem{li}
Kehuang Li and Chin-Hui Lee,
\newblock ``A deep neural network approach to speech bandwidth expansion,''
\newblock in {\em 2015 IEEE International Conference on Acoustics, Speech and
  Signal Processing (ICASSP)}, 2015, pp. 4395--4399.

\bibitem{kuleshov}
Volodymyr Kuleshov et~al.,
\newblock ``Audio super-resolution using neural networks,''
\newblock in {\em ICLR (Workshop)}, 2017.

\bibitem{seanet}
Yunpeng Li et~al.,
\newblock ``Real-time speech frequency bandwidth extension,''
\newblock in {\em ICASSP 2021 - 2021 IEEE International Conference on
  Acoustics, Speech and Signal Processing (ICASSP)}, 2021.

\bibitem{mugan}
Sung Kim and Visvesh Sathe,
\newblock ``Bandwidth extension on raw audio via generative adversarial
  networks,''
\newblock 2019.

\bibitem{bwe-all-you-need}
Jiaqi Su et~al.,
\newblock ``Bandwidth extension is all you need,''
\newblock in {\em ICASSP 2021 - 2021 IEEE International Conference on
  Acoustics, Speech and Signal Processing (ICASSP)}, 2021, pp. 696--700.

\bibitem{nuwave2}
Seungu Han and Junhyeok Lee,
\newblock ``{NU-Wave 2: A General Neural Audio Upsampling Model for Various
  Sampling Rates},''
\newblock in {\em Proc. Interspeech 2022}, 2022, pp. 4401--4405.

\bibitem{wsrglow}
Kexun Zhang, Yi~Ren, Changliang Xu, and Zhou Zhao,
\newblock ``{WSRGlow: A Glow-Based Waveform Generative Model for Audio
  Super-Resolution},''
\newblock in {\em Proc. Interspeech 2021}, 2021, pp. 1649--1653.

\bibitem{lsf}
Sen Li et~al.,
\newblock ``Speech bandwidth extension using generative adversarial networks,''
\newblock in {\em 2018 IEEE International Conference on Acoustics, Speech and
  Signal Processing (ICASSP)}, 2018.

\bibitem{nugan}
Rithesh Kumar et~al.,
\newblock ``Nu-gan: High resolution neural upsampling with gan,''
\newblock 2020.

\bibitem{eskimez}
Sefik~Emre Eskimez and Kazuhito Koishida,
\newblock ``Speech super resolution generative adversarial network,''
\newblock in {\em IEEE International Conference on Acoustics, Speech and Signal
  Processing (ICASSP)}, 2019.

\bibitem{phase-aware}
Shichao Hu et~al.,
\newblock ``{Phase-Aware Music Super-Resolution Using Generative Adversarial
  Networks},''
\newblock in {\em Proc. Interspeech 2020}, 2020, pp. 4074--4078.

\bibitem{hdemucs}
Alexandre D{\'e}fossez,
\newblock ``Hybrid spectrogram and waveform source separation,''
\newblock in {\em Proceedings of the ISMIR 2021 Workshop on Music Source
  Separation}, 2021.

\bibitem{defossez2020real}
Alexandre D{\'e}fossez et~al.,
\newblock ``Real time speech enhancement in the waveform domain,''
\newblock in {\em Interspeech 2020}. 2020, ISCA.

\bibitem{audioPrior}
Arnon Turetzky et~al.,
\newblock ``Deep audio waveform prior,''
\newblock in {\em Interspeech 2022}. 2022, ISCA.

\bibitem{behmgan}
Eloi Moliner and Vesa Välimäki,
\newblock ``Behm-gan: Bandwidth extension of historical music using generative
  adversarial networks,''
\newblock 2022.

\bibitem{choi}
Woosung Choi et~al.,
\newblock ``Investigating u-nets with various intermediate blocks for
  spectrogram-based singing voice separation.,''
\newblock in {\em 21th International Society for Music Information Retrieval
  Conference}, ISMIR, Ed., 2020.

\bibitem{phasen}
Dacheng Yin et~al.,
\newblock ``Phasen: A phase-and-harmonics-aware speech enhancement network,''
\newblock {\em Proceedings of the AAAI Conference on Artificial Intelligence},
  vol. 34, no. 05, pp. 9458--9465, Apr. 2020.

\bibitem{snake}
Liu Ziyin et~al.,
\newblock ``Neural networks fail to learn periodic functions and how to fix
  it,''
\newblock in {\em Advances in Neural Information Processing Systems},
  H.~Larochelle, M.~Ranzato, R.~Hadsell, M.F. Balcan, and H.~Lin, Eds. 2020,
  vol.~33, pp. 1583--1594, Curran Associates, Inc.

\bibitem{yamamoto}
Ryuichi Yamamoto et~al.,
\newblock ``Parallel wavegan: A fast waveform generation model based on
  generative adversarial networks with multi-resolution spectrogram,''
\newblock in {\em ICASSP 2020 - 2020 IEEE International Conference on
  Acoustics, Speech and Signal Processing (ICASSP)}, 2020, pp. 6199--6203.

\bibitem{tfilm}
Sawyer Birnbaum et~al.,
\newblock ``Temporal film: Capturing long-range sequence dependencies with
  feature-wise modulations.,''
\newblock in {\em Advances in Neural Information Processing Systems}. 2019,
  Curran Associates, Inc.

\bibitem{vctk}
Junichi Yamagishi et~al.,
\newblock ``Cstr vctk corpus: English multi-speaker corpus for cstr voice
  cloning toolkit,'' 2019.

\bibitem{musdb18-hq}
Zafar Rafii et~al.,
\newblock ``Musdb18-hq - an uncompressed version of musdb18,'' 2019.

\bibitem{visqol}
Andrew Hines et~al.,
\newblock ``Visqol: an objective speech quality model,''
\newblock {\em EURASIP Journal on Audio, Speech, and Music Processing}, vol.
  2015, no. 1, pp. 1--18, 2015.

\bibitem{webmushra}
Michael Schoeffler et~al.,
\newblock ``webmushra — a comprehensive framework for web-based listening
  tests,''
\newblock {\em Journal of Open Research Software}, vol. 6, no. 1, pp. 8, 2018.

\bibitem{mushra}
ITU,
\newblock ``Itu-r rec. bs.1534-3: Subjective assessment of sound quality,''
  2015.

\end{thebibliography}

\end{document}